\documentclass[prl,a4paper,twocolumn,superscriptaddress,longbibliography]{revtex4-2}

\usepackage{amssymb}
\usepackage{amsmath}
\usepackage{amsfonts}
\usepackage{graphicx}
\usepackage{bm}
\usepackage[usenames,dvipsnames]{xcolor}
\usepackage{multirow}
\usepackage{comment}
\usepackage{pgfplots}
\usepackage{tikz}

\usepackage[colorlinks,citecolor=blue,linkcolor=red,urlcolor=blue]
{hyperref}
\begin{document}

\title{Minimal velocity of the travelling wave solutions in two coupled FKPP equations with the global conservation law}

\author{O. I.  Baburin}

\affiliation{Independent researcher}
\email{baburinoleg740@gmail.com}

\author{I. S. Burmistrov}	

\affiliation{\mbox{L. D. Landau Institute for Theoretical Physics, Semenova 1-a, 142432, Chernogolovka, Russia}}

\affiliation{Laboratory for Condensed Matter Physics, HSE University, 101000 Moscow, Russia}

\date{v1 - \today}

\begin{abstract}
We investigate the system of two coupled one-dimensional Fisher-Kolmogorov-Petrovsky-Piskunov (FKPP) equations which possess the global conservation law. Such system of equations has been recently derived for the quasiparticle densities in the two-band fermionic model with the particle-number conserving dissipative protocol. As standard FKPP equation the studied system of equations has one unstable and one stable homogeneous solution with travelling wave switching between them. We demonstrate that the conservation law enforces the synchronization of travelling waves for both densities and determine their minimal possible velocity. Surprisingly, we find the existence of jumps of the minimal velocity as function of control parameters. We obtain that the minimal velocity of the coupled FKPP equations  may significantly exceed the minimal velocity for a single FKPP equation in a wide range of control parameters. 
\end{abstract}

\maketitle

The Fisher-Kolmogorov-Petrovsky-Piskunov (FKPP) equation \cite{Fisher1937,Kolmogorov1937} is a well-known example belonging to reaction-diffusion universality \cite{FKPP-book,Lutscher2019}. Such type of dynamics emerges in vast majority of  applications, e.g. the propagation of advantageous genes, combustion fronts, the dynamics of domain walls, fluid motion, chemical reactions, decoherence propagation, etc. \cite{FKPP1988,FKPP2,FKPP3,Aleiner2016,Zhou2023}. 

Recently \cite{Nosov2023,Lyublinskaya2025}, the specific form of two coupled FKPP equations has been derived for description of the quasiparticle density evolution in the two-band fermionic model \cite{Tonielli2020,Lyublinskaya2023} with the particle-number conserving dissipative protocol. Such dissipative evolution with the particle conservation
has generated an interest \cite{Sieberer2025} as a way to stabilize topologically nontrivial steady states. 
The particle conservation in the dissipative protocol makes the total number of particles to be the integral of motion of the corresponding FKPP equations. 

The coupled reaction-diffusion equations emerge in many fields, e.g. in physics \cite{Yakupov2019}, in biology \cite{Estavoyer2024,Berestycki2025}, genetics \cite{Ortiz2014}, ecology \cite{Fu2015,Yadav2025aa} as well as they are studied from pure mathematical interest \cite{HOLZER20141,Faye2019}. However, we are not aware about a discussion regarding the role of the global conservation law (the integral of motion) on the behavior of their solutions. 

In this Letter, we report analysis of the two coupled one-dimensional FKPP equations which possess the global conservation law. Similar to a standard FKPP equation, the studied system of equations, cf. Eq. \eqref{eq:FKPP}, has an unstable and stable homogeneous solutions with travelling waves switching between them. We demonstrate that the conservation law enforces the synchronization of travelling waves for both densities and determine their minimal possible velocity. Surprisingly, we find the existence of jumps of the minimal velocity as function of control parameters. We obtain that the minimal velocity of the coupled FKPP equations may significantly exceed the minimal velocity for a single FKPP equation in a wide range of control parameters. 

\noindent\textsf{\color{blue} Formulation of the problem. ---} We study the following two coupled 1D FKPP equations \cite{Lyublinskaya2025}
\begin{equation}
\begin{aligned}
\partial_t n_u - D_u \partial^2_x n_u - \frac{n_u}{\tau_u} + \frac{n_d}{\tau_d} - \beta n_u  n_d = 0 , \\
\partial_t n_d - D_d \partial^2_x  n_d - \frac{n_d}{\tau_d}+ \frac{n_u}{\tau_u} + \beta n_u n_d = 0 .
\end{aligned}
\label{eq:FKPP}
\end{equation}
Here $n_u(x,t){\geqslant}0$ describes the density of particles in the upper band, while $n_d(x,t){\leqslant} 0$ is the density of holes in the lower band. Physically, $n_{u,d}$ are small and corresponds to the deviations of the densities from the half-filled case (the upper band is empty while the lower band is fully occupied): $n_u{=}n_d{=}0$. 
$D_{u,d}{>}0$ denote the diffusion coefficients in each band. 
The rates $1/\tau_{u,d}{>}0$ characterize instability of the fully occupied lower band towards the process similar to the impact ionization in semiconductors \cite{Keldysh1960}. The quantity $\beta{>}0$ describes the rate of the recombination akin the non-radiatve recombination in semiconductors \cite{Abakumov-book}. We note that Eq.~\eqref{eq:FKPP} seems to be generic and might describe the two types of diffusive particles with recombination and interparticle transitions.

The system \eqref{eq:FKPP} has unstable stationary homogeneous solution $n_u{=}n_d{=}0$ and stable stationary homogeneous solution $n_u{=}{-}n_d{=}n_*{=}\Gamma/\beta$, where $\Gamma{=}1/\tau_u{+}1/\tau_d$. Also the dynamics \eqref{eq:FKPP} has the global conserved quantity, the total number of excess particles $N$:
\begin{equation}
  N  = \int dx \Bigl [n_u(x,t) + n_d(x,t)\Bigr ] = 0 .
 \label{eq:conserv}
\end{equation}
The above constraints holds under natural assumptions that $\lim\limits_{x{\to}\infty}\partial_x n_a(x,t){=}0$.
We emphasize that the conservation of the total number of particles impose strong constraint on the dynamics of the system \eqref{eq:FKPP}. Indeed, it makes that the homogeneous solutions of Eq. \eqref{eq:FKPP} for $n_u$ and $n_d$ to be not independent: 
\begin{equation}
 n_u(t)=-n_d(t) = \frac{n(0) n_* e^{\Gamma t}}{n_*+n(0)(e^{\Gamma t}-1)} .
 \label{eq:Sol:Hom}
\end{equation}

It is convenient to introduce the following dimensionless time $\tau$ and distance $\rho$: 
$t{=}(\tau_u\tau_d)^{1/2}\tau$ and  $x{=}(D_u D_d \tau_u\tau_d)^{1/4} \rho$. Also we introduce dimensionless densities $U{=}n_u \beta\sqrt{\tau_u\tau_d}$ and $V{=}n_d \beta\sqrt{\tau_u\tau_d}$.
Then Eqs.~\eqref{eq:FKPP} become
\begin{equation}
\begin{aligned}
\partial_\tau U - \frac{1}{\sqrt{\alpha}} \partial^2_\rho U - \frac{1}{\sqrt{\delta}} U+  \sqrt{\delta} V - U V = 0 , \\
\partial_\tau V - \sqrt{\alpha} \partial^2_\rho V -  \sqrt{\delta} V + \frac{1}{\sqrt{\delta}} U + U V = 0  .
\end{aligned}
\label{eq:FKPP:1}
\end{equation}
Here  we introduced two control parameters $\alpha{=}D_d/D_u$ and $\delta{=}\tau_u/\tau_d$. In dimensional units the unstable solution is $U{=}V{=}0$ while the stationary one corresponds to $U{=}{-}V{=}U_*{=}\sqrt{\delta}{+}1/\sqrt\delta$. Also we note that the dynamics of $U$ and $V$ is such that their magnitudes belong to the region $\sqrt{\delta}/U{+}1/(|V|\sqrt{\delta}){\geqslant} 1$.

\noindent\textsf{\color{blue} Synchronization of the travelling wave solutions. ---} As for the single FKPP equation, we are interested in the solutions of Eqs. \eqref{eq:FKPP:1} that satisfy the following initial conditions
\begin{equation}
\begin{aligned}
\lim\limits_{\rho\to-\infty} U(\rho,0) & =- \lim\limits_{\rho\to-\infty} V(\rho,0) =\sqrt{\delta}{+}1/\sqrt\delta  , \\
\lim\limits_{\rho\to+\infty} U(\rho,0) & = \lim\limits_{\rho\to+\infty} V(\rho,0) = 0 .    
\end{aligned}
\label{eq:UV:tt}
\end{equation}
We seek for the traveling wave solutions: 
$U(\rho,\tau){=}\mathcal{U}(\rho{-}v_1 \tau)$ and $V(\rho,t){=}\mathcal{V}(\rho{-}v_2 \tau)$ with two independent velocities $v_1$ and $v_2$, in general. However, we find
\begin{equation}
\partial_t N \propto -\int d\rho  \Bigl [v_1 \mathcal{U}^\prime (\rho-v_1 \tau)+v_2 \mathcal{V}^\prime(\rho-v_2 \tau)\Bigr ] = 0 ,
\end{equation}
provided $v_1{=}v_2$. Therefore, the conservation law \eqref{eq:conserv} forces synchronization of the travelling wave solutions making their velocities identical, $v_1{=}v_2{=}v$. Below we assume that $v{>}0$. The results for $v{<}0$ can be obtained by means of trivial redefinitions.

\noindent\textsf{\color{blue} Asymptotic analysis near the unstable solution. ---} Substituting the travelling wave ansatz $U(\rho,\tau){=}\mathcal{U}(z)$ and $V(\rho,\tau){=}\mathcal{V}(z)$, where $z{=}\rho{-}v\tau$, into Eqs.~\eqref{eq:FKPP:1} and taking the limit $z{\to}{+}\infty$ (for which $\mathcal{U},\mathcal{V}{\to} 0$, we obtain the linearized system of equations
\begin{equation}
 \begin{pmatrix}
 v \partial_z +\frac{1}{\sqrt{\alpha}} \partial^2_z+ \frac{1}{\sqrt{\delta}} &   -\sqrt{\delta} \\
  -\frac{1}{\sqrt{\delta}} & v \partial_z + \sqrt{\alpha} \partial^2_z+ \sqrt{\delta} 
 \end{pmatrix}
  \begin{pmatrix}
    \mathcal{U} \\
    \mathcal{V}
 \end{pmatrix} = 0 .
\end{equation}
Seeking the solution in the standard form $\mathcal{U},\mathcal{V}{\propto} \exp(\lambda z)$, we obtain the characteristic equation:
\begin{equation}
\lambda^4{+}2 v \lambda^3 \cosh a
{+} [v^2{+}2 \cosh(a{-}b)] \lambda^2{+}2 v \lambda\cosh b {=} 0 .
\label{eq:lambda}
\end{equation}
Here we parametrize $\alpha$ and $\delta$ as $\alpha{=}\exp(2a)$ and $\delta{=}\exp(2b)$. We are interested in real negative solutions of Eq. \eqref{eq:lambda} in order to preserve the non-negativity of $\mathcal{U}$ and ${-}\mathcal{V}$, as well as to guarantee that $\mathcal{U},\mathcal{V}{\to} 0$ as $z{\to}{+}\infty$. 
Due to inequality $2 v \cosh b {>}0$, Eq. \eqref{eq:lambda} has at least one real negative root. 

Let us first consider the case $\alpha{=}1$ ($a{=}0$). Then we obtain the following four roots of Eq.~\eqref{eq:lambda}:
\begin{equation}
\lambda_1=0, \, \lambda_2=-v, \quad \lambda_{3,4} = -\frac{v \pm \sqrt{v^2 - 8\cosh b}}{2} .
\end{equation}
The roots $\lambda_{1,2}$ controls asymptotic behavior of the total density $\mathcal{U}{+}\mathcal{V}$. The existence of zero root, $\lambda_1{=}0$, is the consequence of the conservation law \eqref{eq:conserv}
for the total number of particles. The other two roots, $\lambda_{3,4}$ corresponds to the eigenvector $\{1,-1\}^T$ and describes the asymptotic behavior of the imbalance $\mathcal{U}{-}\mathcal{V}$. For $v{\to}\infty$ the root $\lambda_{4}$ tends to the magnitude $2v \cosh b$ that corresponds to the homogeneous solution \eqref{eq:Sol:Hom}. 

In order the roots $\lambda_{3,4}$ to be real, velocity has to be larger then the minimal one: $v^2 {\geqslant} 8\cosh b$. 
Returning to the dimensional units for the velocity: $c{=}v (D_u D_d/(\tau_u\tau_d))^{1/4}$, we find that for the case of equal diffusion coefficients $D_u{=}D_d$ ($\alpha{=}1$) thetravelling wave velocity satisfies the following inequality: $c^2{\geqslant}c_{\rm min}^2{=}c_{\rm min, u}^2{+}c_{\rm min, d}^2$ where $c_{\rm min, a}{=}2\sqrt{D_a/\tau_a}$ is the minimal velocity for a single FKPP equation.

\begin{figure}[t]
\centerline{\includegraphics[width=0.75\columnwidth]{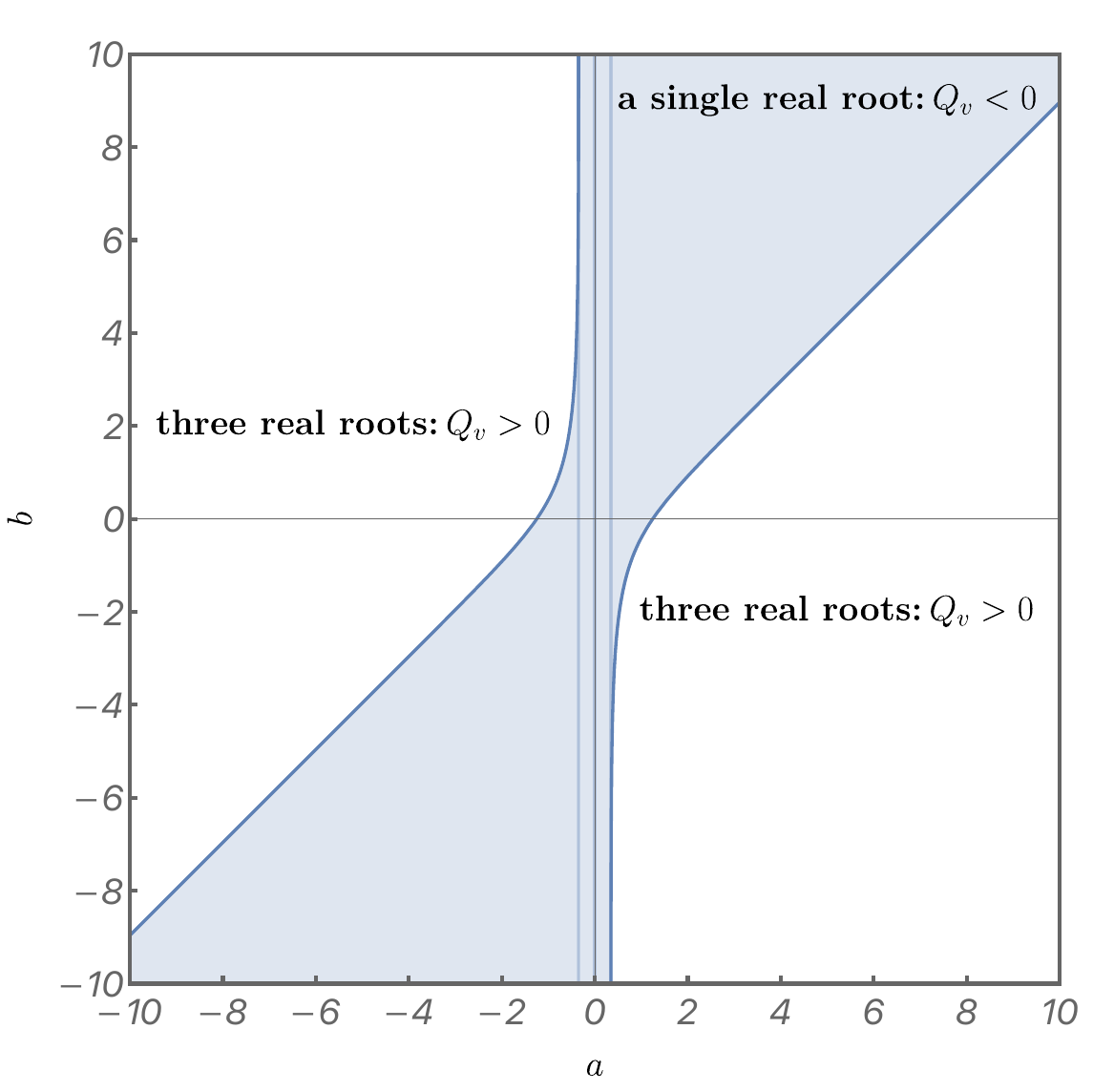}}
    \caption{The regions of positive and negative $Q_v$ on the $b-a$ plane, see Eq. \eqref{eq:p1}.}
    \label{fig-p1}
\end{figure}

Now we return to the general case. The discriminant of the qubic equation corresponding to Eq.~\eqref{eq:lambda} acquires the following form:
\begin{widetext}
\begin{gather}
Q_\lambda {=} {-} v^6 \sinh^2a {+} 2 v^4 \sinh a  [4 \cosh(a {+} b) \sinh a {+} \sinh(2 a {-} b)] 
{+}
v^2 [4(1 {-} 2 \cosh(2 a)) \cosh^2(a{-}b) \notag \\ {+} 
    3 (\cosh(2 a {-} b) {-} 2 \cosh b )^2] {+} 8 \cosh^3(a {-} b) .
    \label{eq:Ql}
   \end{gather} 
\end{widetext}
We note that $Q_\lambda$ is not changed after simultaneous change of the signs of $a$ and $b$: $a,b\to{-}a,{-}b$.
For $Q_\lambda{<}0$ there exists a single real negative root of Eq. \eqref{eq:lambda} and two compex conjugated roots. Such situation does not allow us to construct asymptotic solution for $\mathcal{U}{\geqslant} 0$ and $\mathcal{V}{\leqslant}0$. In the opposite case $Q_\lambda{>}0$ we have three real roots with at least one negative one. With three such roots we can build the proper asymptotic behavior. The minimal velocity required to have three real roots is determined from the equation  $Q_\lambda{=}0$. The corresponding cubic equation for $v^2$ 
has at least one positive real root. Again in order to characterize the number of real roots for $v^2$ we compute the discriminant of the corresponding qubic equation:
\begin{gather}
Q_v {=} \cosh b\Bigl [(12 {-} 14 \cosh(2 a) {+} \cosh(4 a)) \cosh b \notag \\ {-} 
   8 \cosh^3 a \sinh a  \sinh b \Bigr ]^3 .
   \label{eq:Qv}
\end{gather}
For $Q_v{<}0$ there is only single real positive root for $v^2$. In the opposite case of $Q_v{>}0$, there are three real roots. Thus the minimal positive root determines the minimal velocity square in this case. The boundary between two regimes are determined by the equation $Q_v{=}0$. It can be resolved explicitly in terms of dependence of $b$ on $a$:
\begin{equation}
b = a+ \frac{3}{2} \ln \frac{3 \tanh a-1}{3\tanh a+1} . 
\label{eq:p1}
\end{equation}
We note that for fixed $b$ there are two solutions of the above equation: $a_b^\pm$ with $a_b^- {<} 0 {<} a_b^+$. For $a{<}a_b^-$ we have three real solutions for $v^2$, $Q_v{>}0$. For $a_b^-{<}a{<}a_b^+$ there is only a single real solution. Finally, for $a_b^+{<}a$ we have three real solutions again. We note that $|a_b^{\pm}|{>}(1/2)\ln 2$ (see Fig. \ref{fig-p1}). 

%%%%%%%%%%%%%%%%%%%%%%%%%%%%%%%%%%%%%%%%
%%%%%%%%%%%%%%%%%%%%%%%%%%%%%%%%%%%%%%%%%

\begin{figure}[t]
\centerline{\includegraphics[width=0.95\columnwidth]{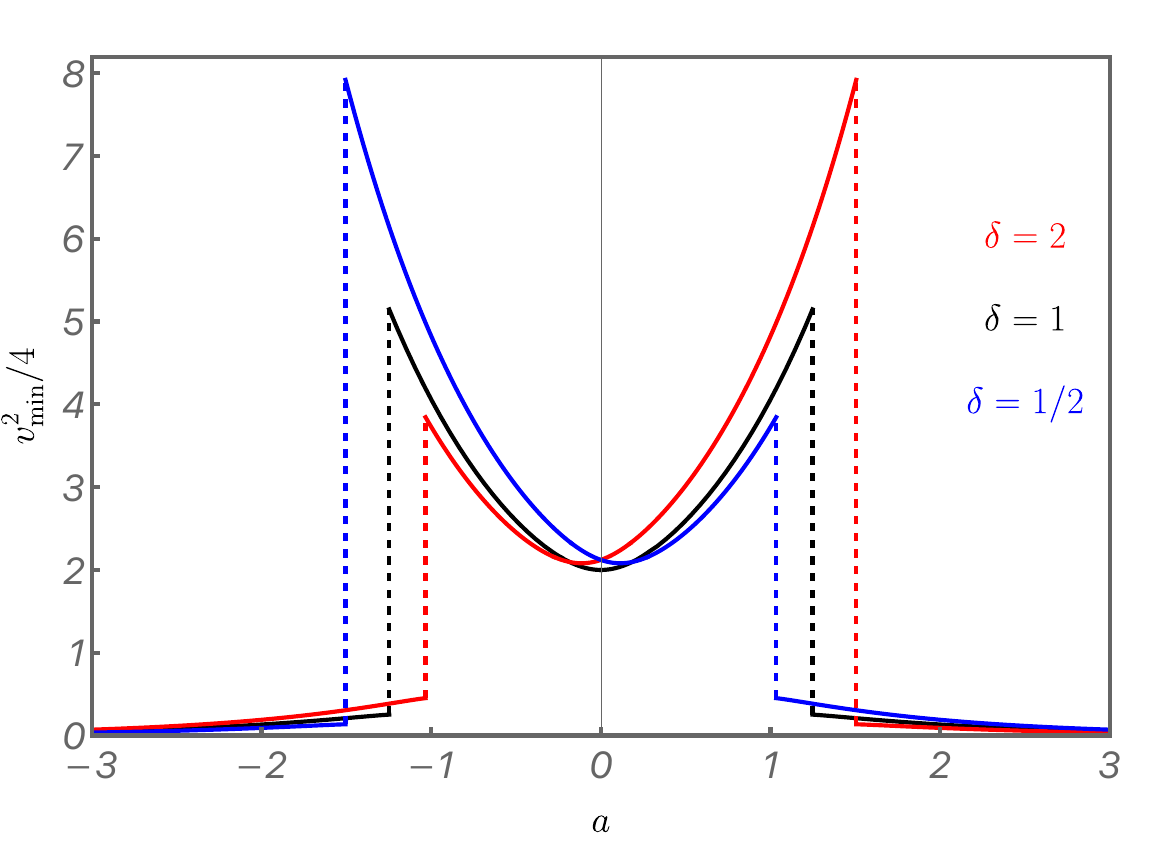}}
    \caption{Dependence of $v^2_{\rm min}$ on $a{=}(1/2)\ln \alpha$ for $\delta{=}1/2, 1, 2$. The jumps occur at the points $a_b^{\pm}$ which are the solutions of Eq. \eqref{eq:p1}.}
    \label{fig-d1}
\end{figure}

 Instead of writing long analytic expressions for the minimal velocity $v$, we present the following asymptotic results 
\begin{equation}
v^2_{\rm min} \simeq 4 \begin{cases}
2 \cosh b {+}2 a \sinh b {+} a^2 \frac{25 \cosh(2 b)-19}{2 \cosh b} , & a{\to} 0 \\
e^{-|a+b|}-e^{-3|a+b|}, &  a{\to}\infty ,
%e^{a+b}(1-e^{2a+2b}), & \quad a\to -\infty
\end{cases}
\label{eq:asym}
\end{equation}
and numerical dependence of $v^2_{\rm min}$ on $a$ for a fixed $b$ (see Fig. \ref{fig-d1}).

%%%%%%%%%%%%%%%%%%%%%%%%%%%%%%%%%%%%%%%%
%%%%%%%%%%%%%%%%%%%%%%%%%%%%%%%%%%%%%%%%%

\noindent\textsf{\color{blue} Asymptotic analysis near the stable solution. ---} Now let us consider vicinity of the stable homogeneous solution  $U{=}{-}V{=}2\cosh b$. Since this solution is expected to be realized at $z{\to}{-}\infty$, we are interested in the positive eigenvalues $\bar\lambda$, where we write $\mathcal{U}{-}2\cosh b \propto \exp(\bar{\lambda}z)$ and $\mathcal{V}{+}2\cosh b \propto \exp(\bar{\lambda}z)$. Then the characteristic equation for the corresponding eigenvalues becomes
\begin{equation}
\bar{\lambda}^4{+}2 v \bar{\lambda}^3 \cosh a
{+} [v^2{-}2 \cosh(a{-}b)] \bar{\lambda}^2 {-}2 v \bar{\lambda} \cosh b {=} 0 .
\label{eq:lambda:1}
\end{equation}
Since the last term in the right hand side, ${-}2 v \cosh b{<}0$, Eq. \eqref{eq:lambda:1} has at least one real positive root. As one can check the discriminant of Eq. \eqref{eq:lambda:1} coincides with $Q_v$. Therefore, the analysis of the vicinity of the stable stationary solution does not impose additional conditions for $v_{\rm min}$. 

\begin{figure*}[t]
\centerline{\includegraphics[width=0.3\textwidth]{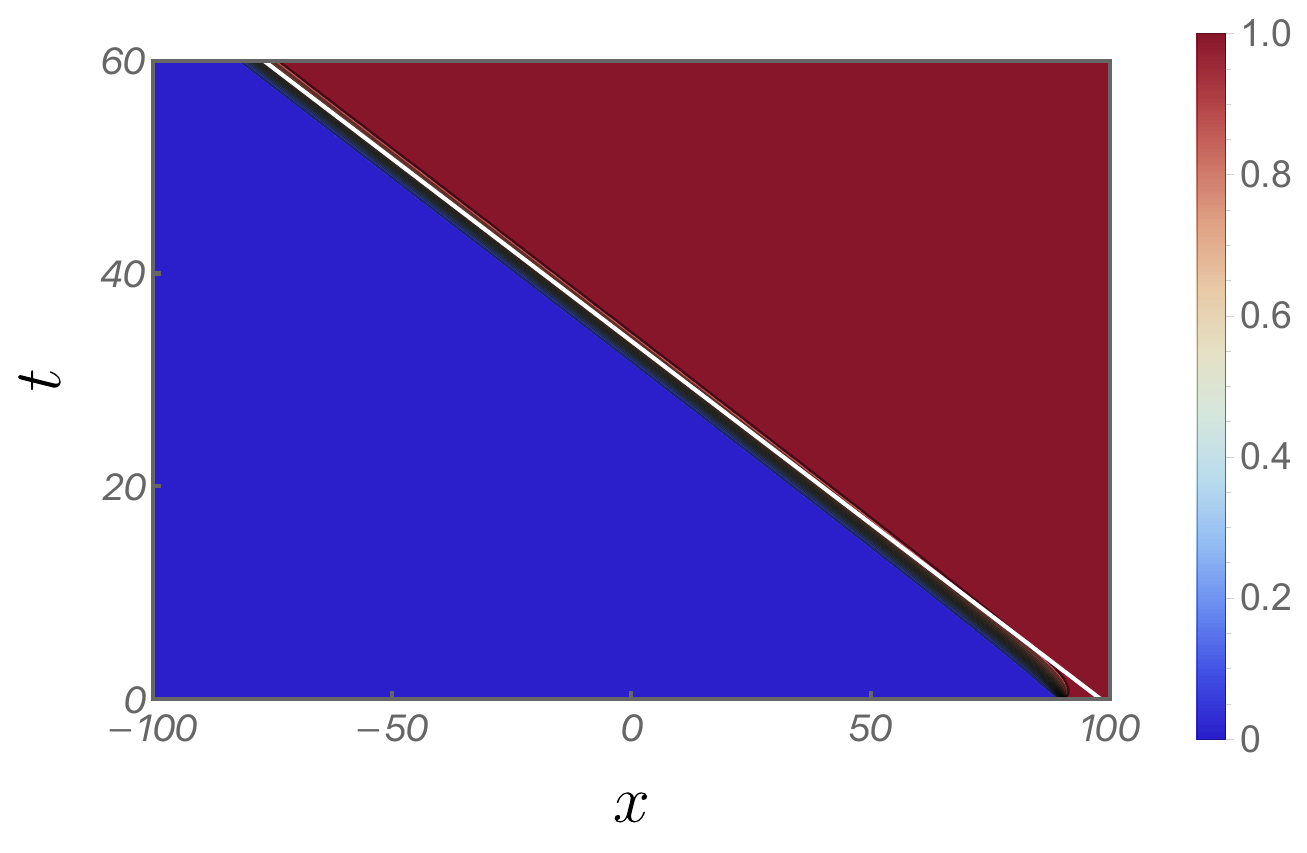}\qquad \includegraphics[width=0.3\textwidth]{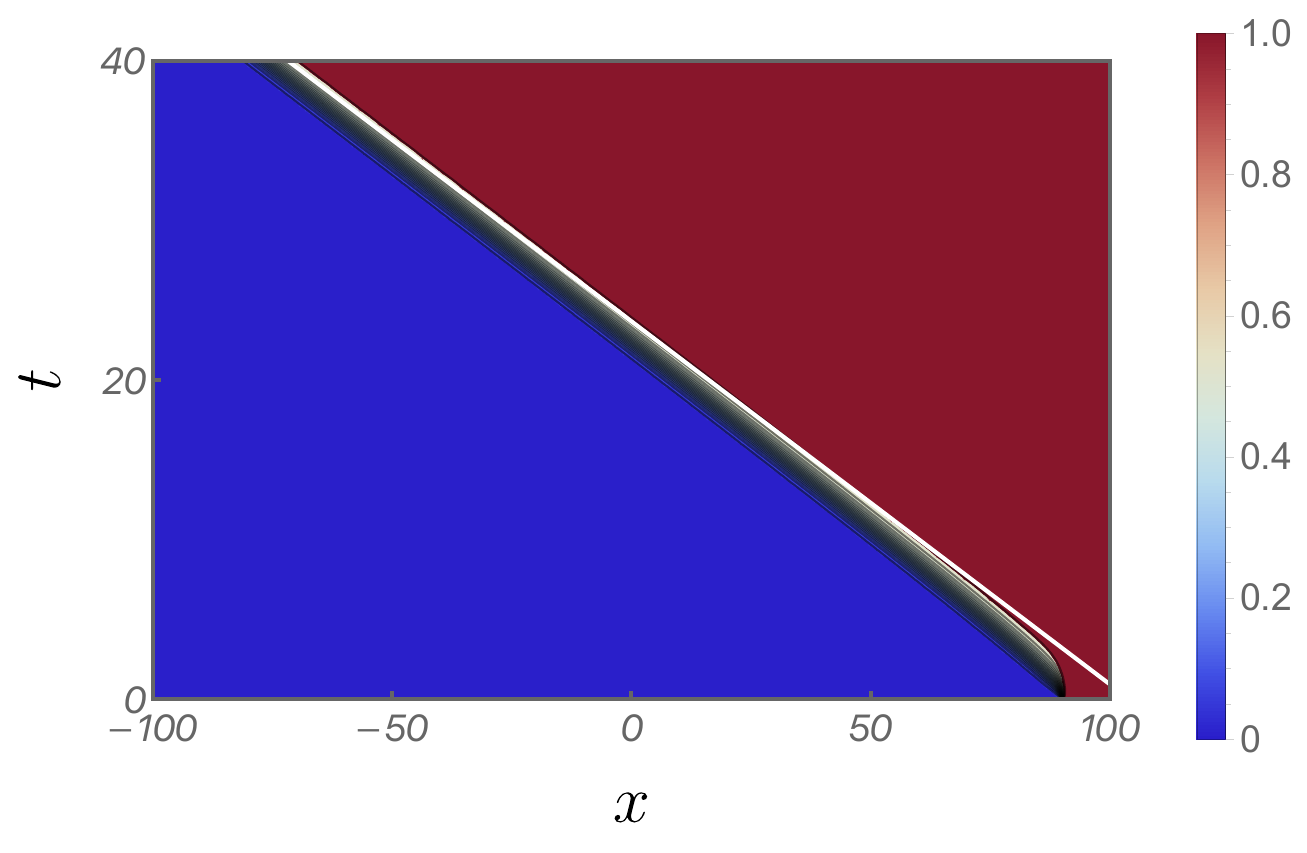}\qquad \includegraphics[width=0.3\textwidth]{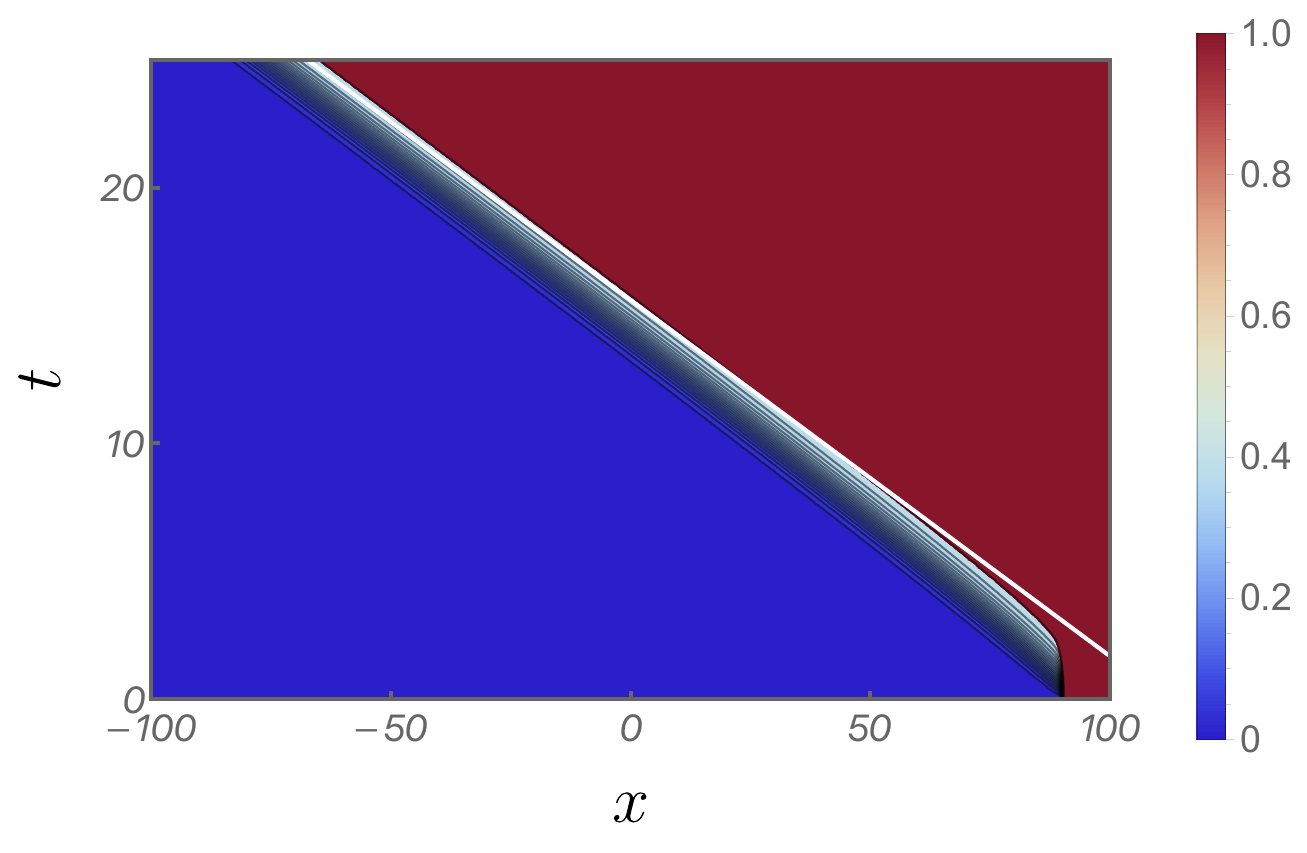}}
\centerline{\includegraphics[width=0.3\textwidth]{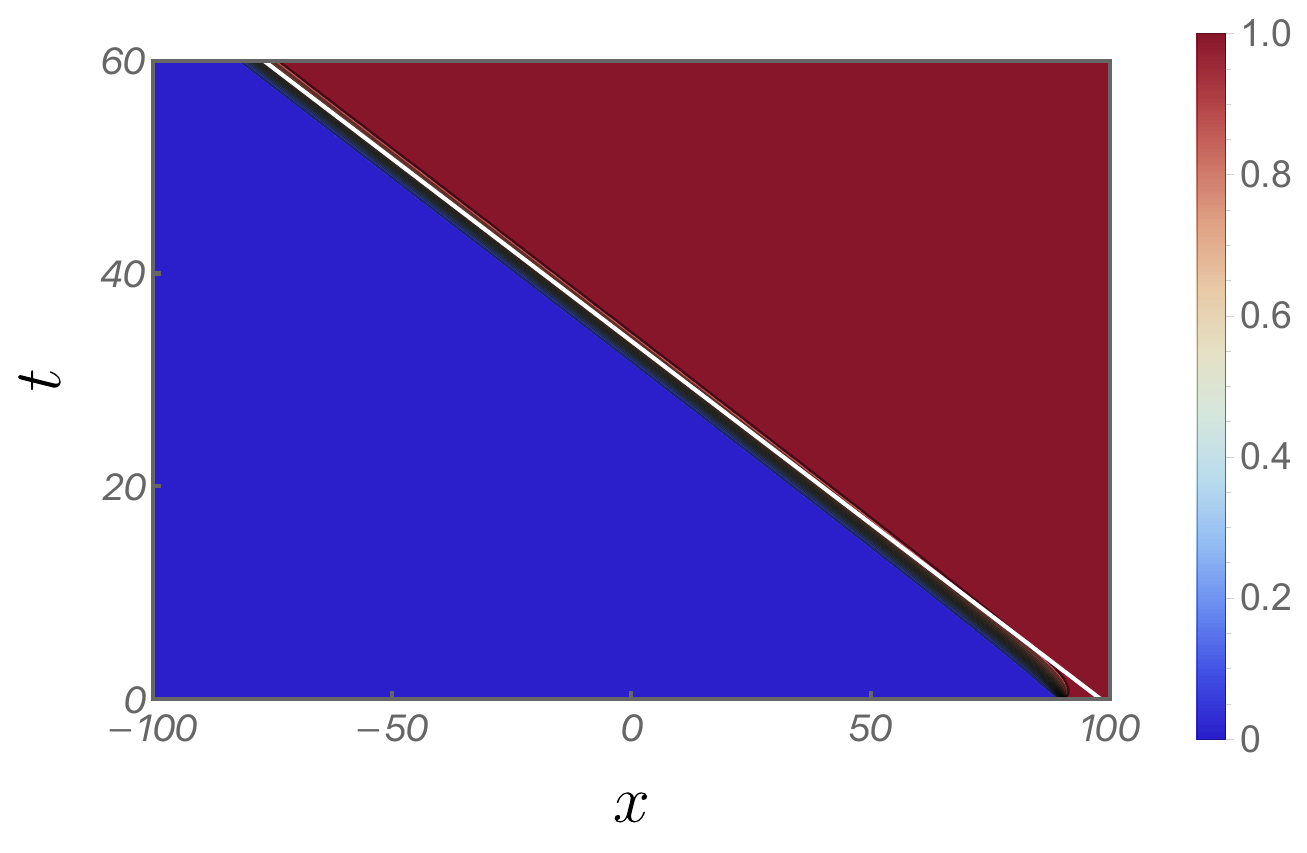}\qquad \includegraphics[width=0.3\textwidth]{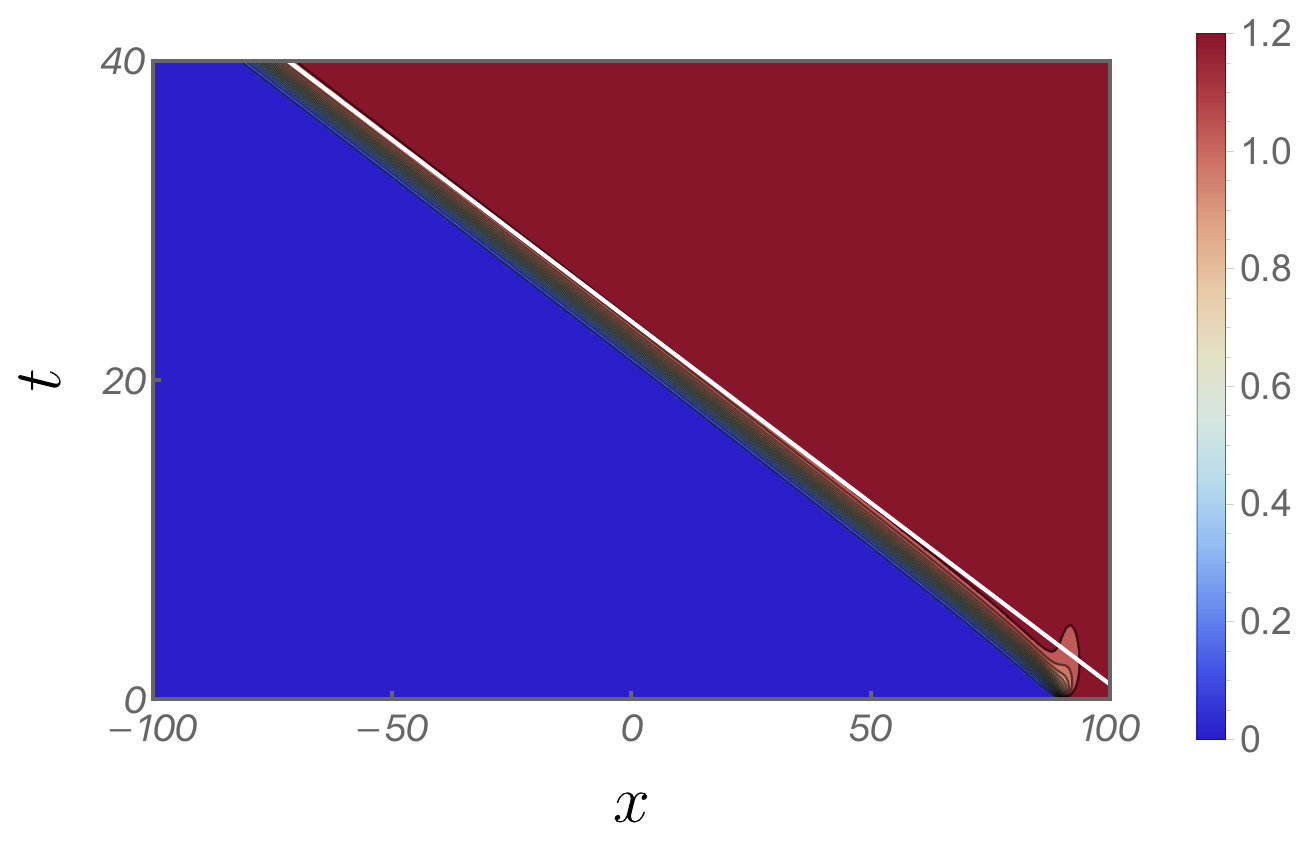}\qquad \includegraphics[width=0.3\textwidth]{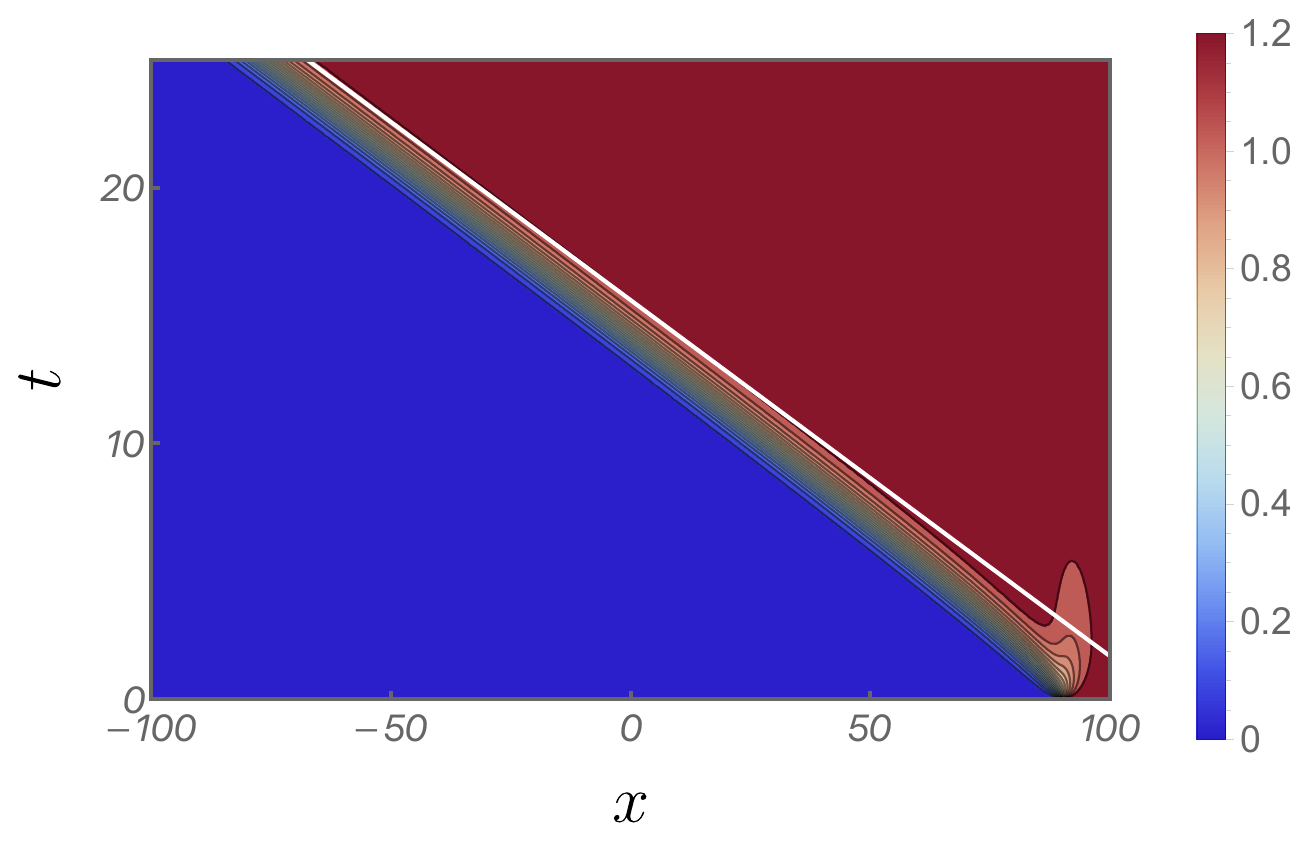}}
    \caption{The color-density plot of $U(t,x)/U_*$ (upper row) and ${-}V(t,x)/U*$ (lower row) for $\delta{=}2$ and $a{=}0$ (left column), $a{=}1$ (middle column) and $a{=}2$ (right column). We solve Eqs. \eqref{eq:FKPP:1} at the interval $x{\in} [{-}100,100]$ with the following initial and boundary conditions: 
    $U(0,x){=}{-}V(0,x){=}U_*\theta(x{-}90)$, $U(t,{-}100){=}V(t,{-}100){=}U_*$, $U(t,100){=}V(t,100){=}U_*$. 
    The straight white line corresponds to the velocity of the travelling wave, $v\simeq 2.91, 4.39, 7.18$, for left, middle, and right columns respectively. }
    \label{fig-num}
\end{figure*}

\noindent\textsf{\color{blue} Numerical results. --- } To illustrate our analytical results we solve Eqs. \eqref{eq:FKPP:1} numerically using sharp initial conditions $U(0,x){=}{-}V(0,x){=}U_*\theta(x-x_0)$, where $\theta(x)$ denotes the Heaviside step function. The numerical results for $\delta{=}2$ and for three values $a{=}0,1,2$  are shown in Fig.~\ref{fig-num}. We demonstrate that the velocities of travelling waves for $U$ and $V$ are indeed synchronized. Interestingly, for such sharp initial conditions the velocity of the travelling waves in the case of $a{=}0$ and $a{=}1$ coincides exactly with the minimal velocity for a given value of $a$ (which is smaller than $a_b^+$. In contrast, the velocity for $a{=}2$ is much larger than the minimal velocity, which is equal to $0.61$. In fact the observed velocity for $a{=}2$ coincides with maximal real root of the equation $Q_v{=}0$, see Eq.~\eqref{eq:Qv}.

%%%%%%%%%%%%%%%%%%%%%%%%%%%%%%%%%%%%%%%%
%%%%%%%%%%%%%%%%%%%%%%%%%%%%%%%%%%%%%%%%%

\noindent\textsf{\color{blue} Conclusions. --- } In our study we investigated the behavior of the system of two coupled one-dimensional FKPP equations for densities of fermions in the upper and lower bands. The crucial difference of the considered system of equations from the ones studied before is the existence of the global conservation law. The phenomenology for the behavior of 
the studied system of equation is similar to the standard FKPP equation: the existence of unstable and stable stationary homogeneous solutions as well as travelling waves switching the system between them. However, 
the conservation law enforces the synchronization of travelling waves for both densities and determine their minimal possible velocity. The conservation law results in the existence of zero root of the characteristic polynomial for equations linearized near the unstable and stable solutions. We performed the linear analysis and found analytically the minimal velocity at which the travelling waves can spread out. We obtained that the minimal velocity is a non-trivial function of the two control parameters of the model. Moreover, we demonstrated the existence of jumps of the minimal velocity as function of these control parameters. Interestingly, we found that the minimal velocity of the coupled FKPP equations may significantly exceed the minimal velocity for a single FKPP equation in a wide range of parameters. 

As a possible direction for future research we mention a study of possibility of  existence of nontrivial spatial structures at the front of the travelling wave in the case of two and three dimensions (see e.g. \cite{Yakupov2019}). Also we note that for the ordinary one-dimensional FKPP equation the exact solution for the travelling wave with a particular velocity is known \cite{Ablowitz1979}. It would be interesting to find similar exact solutions for the coupled FKPP equations, perhaps, using the approach of Ref. \cite{Manakov1987}. In addition, it would be worthwhile to extend our analysis to more general form of equations of FKPP type which possesses the global conservation law.

\noindent\textsf{\color{blue} Acknowledgements. ---} We thank E.A. Kuznetsov for the interest to this problem. 
I.S.B. is grateful to A.A. Lyublinskaya and P.A. Nosov for collaboration on the related project. The work of I.S.B. was supported by the Russian Ministry of Science and Higher Education and by the Basic Research Program of HSE. I.S.B. acknowledges personal support from the Foundation for the Advancement of Theoretical Physics and Mathematics “BASIS”. O.I.B. acknowledges the hospitality at Landau Institute for Theoretical Physics during the Summer School on Theoretical Physics 2025 where this work has been performed.

\bibliography{Lindblad-ref,FKPP-ref}

%apsrev4-2.bst 2019-01-14 (MD) hand-edited version of apsrev4-1.bst
%Control: key (0)
%Control: author (8) initials jnrlst
%Control: editor formatted (1) identically to author
%Control: production of article title (0) allowed
%Control: page (0) single
%Control: year (1) truncated
%Control: production of eprint (0) enabled
\begin{thebibliography}{26}%
\makeatletter
\providecommand \@ifxundefined [1]{%
 \@ifx{#1\undefined}
}%
\providecommand \@ifnum [1]{%
 \ifnum #1\expandafter \@firstoftwo
 \else \expandafter \@secondoftwo
 \fi
}%
\providecommand \@ifx [1]{%
 \ifx #1\expandafter \@firstoftwo
 \else \expandafter \@secondoftwo
 \fi
}%
\providecommand \natexlab [1]{#1}%
\providecommand \enquote  [1]{``#1''}%
\providecommand \bibnamefont  [1]{#1}%
\providecommand \bibfnamefont [1]{#1}%
\providecommand \citenamefont [1]{#1}%
\providecommand \href@noop [0]{\@secondoftwo}%
\providecommand \href [0]{\begingroup \@sanitize@url \@href}%
\providecommand \@href[1]{\@@startlink{#1}\@@href}%
\providecommand \@@href[1]{\endgroup#1\@@endlink}%
\providecommand \@sanitize@url [0]{\catcode `\\12\catcode `\$12\catcode
  `\&12\catcode `\#12\catcode `\^12\catcode `\_12\catcode `\%12\relax}%
\providecommand \@@startlink[1]{}%
\providecommand \@@endlink[0]{}%
\providecommand \url  [0]{\begingroup\@sanitize@url \@url }%
\providecommand \@url [1]{\endgroup\@href {#1}{\urlprefix }}%
\providecommand \urlprefix  [0]{URL }%
\providecommand \Eprint [0]{\href }%
\providecommand \doibase [0]{https://doi.org/}%
\providecommand \selectlanguage [0]{\@gobble}%
\providecommand \bibinfo  [0]{\@secondoftwo}%
\providecommand \bibfield  [0]{\@secondoftwo}%
\providecommand \translation [1]{[#1]}%
\providecommand \BibitemOpen [0]{}%
\providecommand \bibitemStop [0]{}%
\providecommand \bibitemNoStop [0]{.\EOS\space}%
\providecommand \EOS [0]{\spacefactor3000\relax}%
\providecommand \BibitemShut  [1]{\csname bibitem#1\endcsname}%
\let\auto@bib@innerbib\@empty
%</preamble>
\bibitem [{\citenamefont {Fisher}(1937)}]{Fisher1937}%
  \BibitemOpen
  \bibfield  {author} {\bibinfo {author} {\bibfnamefont {R.~A.}\ \bibnamefont
  {Fisher}},\ }\bibfield  {title} {\bibinfo {title} {The wave of advance of
  advantageous genes},\ }\href@noop {} {\bibfield  {journal} {\bibinfo
  {journal} {Ann. Eugen.}\ }\textbf {\bibinfo {volume} {7}},\ \bibinfo {pages}
  {355} (\bibinfo {year} {1937})}\BibitemShut {NoStop}%
\bibitem [{\citenamefont {Kolmogorov}\ \emph {et~al.}(1937)\citenamefont
  {Kolmogorov}, \citenamefont {Petrovsky},\ and\ \citenamefont
  {Piscounov}}]{Kolmogorov1937}%
  \BibitemOpen
  \bibfield  {author} {\bibinfo {author} {\bibfnamefont {A.}~\bibnamefont
  {Kolmogorov}}, \bibinfo {author} {\bibfnamefont {I.}~\bibnamefont
  {Petrovsky}},\ and\ \bibinfo {author} {\bibfnamefont {N.}~\bibnamefont
  {Piscounov}},\ }\bibfield  {title} {\bibinfo {title} {Study of the diffusion
  equation with growth of the quantity of matter and its application to a
  biological problem},\ }\href@noop {} {\bibfield  {journal} {\bibinfo
  {journal} {Bull. State Univ. Mosc.}\ }\textbf {\bibinfo {volume} {1}},\
  \bibinfo {pages} {1} (\bibinfo {year} {1937})}\BibitemShut {NoStop}%
\bibitem [{\citenamefont {Danilov}\ \emph {et~al.}(1995)\citenamefont
  {Danilov}, \citenamefont {Maslov},\ and\ \citenamefont
  {Volosov}}]{FKPP-book}%
  \BibitemOpen
  \bibfield  {author} {\bibinfo {author} {\bibfnamefont {V.~G.}\ \bibnamefont
  {Danilov}}, \bibinfo {author} {\bibfnamefont {V.~P.}\ \bibnamefont
  {Maslov}},\ and\ \bibinfo {author} {\bibfnamefont {K.~A.}\ \bibnamefont
  {Volosov}},\ }\href@noop {} {\emph {\bibinfo {title} {Mathematical Modelling
  of Heat and Mass Transfer Processes}}}\ (\bibinfo  {publisher} {Springer
  Science and Business Media, B.V.},\ \bibinfo {year} {1995})\BibitemShut
  {NoStop}%
\bibitem [{\citenamefont {Lutscher}(2019)}]{Lutscher2019}%
  \BibitemOpen
  \bibfield  {author} {\bibinfo {author} {\bibfnamefont {F.}~\bibnamefont
  {Lutscher}},\ }\href {https://doi.org/10.1007/978-3-030-29294-2} {\emph
  {\bibinfo {title} {Integrodifference Equations in Spatial Ecology}}}\
  (\bibinfo  {publisher} {Springer International Publishing},\ \bibinfo {year}
  {2019})\BibitemShut {NoStop}%
\bibitem [{\citenamefont {van Saarloos}(1988)}]{FKPP1988}%
  \BibitemOpen
  \bibfield  {author} {\bibinfo {author} {\bibfnamefont {W.}~\bibnamefont {van
  Saarloos}},\ }\bibfield  {title} {\bibinfo {title} {Front propagation into
  unstable states: Marginal stability as a dynamical mechanism for velocity
  selection},\ }\href {https://doi.org/10.1103/PhysRevA.37.211} {\bibfield
  {journal} {\bibinfo  {journal} {Phys. Rev. A}\ }\textbf {\bibinfo {volume}
  {37}},\ \bibinfo {pages} {211} (\bibinfo {year} {1988})}\BibitemShut
  {NoStop}%
\bibitem [{\citenamefont {Brunet}\ and\ \citenamefont {Derrida}(2015)}]{FKPP2}%
  \BibitemOpen
  \bibfield  {author} {\bibinfo {author} {\bibfnamefont {{\'E}.}~\bibnamefont
  {Brunet}}\ and\ \bibinfo {author} {\bibfnamefont {B.}~\bibnamefont
  {Derrida}},\ }\bibfield  {title} {\bibinfo {title} {An exactly solvable
  travelling wave equation in the {Fisher--KPP} class},\ }\href
  {https://doi.org/https://doi.org/10.1007/s10955-015-1350-6} {\bibfield
  {journal} {\bibinfo  {journal} {J. Stat. Phys.}\ }\textbf {\bibinfo {volume}
  {161}},\ \bibinfo {pages} {801} (\bibinfo {year} {2015})}\BibitemShut
  {NoStop}%
\bibitem [{\citenamefont {Derrida}\ and\ \citenamefont {Spohn}(1988)}]{FKPP3}%
  \BibitemOpen
  \bibfield  {author} {\bibinfo {author} {\bibfnamefont {B.}~\bibnamefont
  {Derrida}}\ and\ \bibinfo {author} {\bibfnamefont {H.}~\bibnamefont
  {Spohn}},\ }\bibfield  {title} {\bibinfo {title} {Polymers on disordered
  trees, spin glasses, and traveling waves},\ }\href
  {https://doi.org/https://doi.org/10.1007/BF01014886} {\bibfield  {journal}
  {\bibinfo  {journal} {J. Stat. Phys.}\ }\textbf {\bibinfo {volume} {51}},\
  \bibinfo {pages} {817} (\bibinfo {year} {1988})}\BibitemShut {NoStop}%
\bibitem [{\citenamefont {Aleiner}\ \emph {et~al.}(2016)\citenamefont
  {Aleiner}, \citenamefont {Faoro},\ and\ \citenamefont {Ioffe}}]{Aleiner2016}%
  \BibitemOpen
  \bibfield  {author} {\bibinfo {author} {\bibfnamefont {I.~L.}\ \bibnamefont
  {Aleiner}}, \bibinfo {author} {\bibfnamefont {L.}~\bibnamefont {Faoro}},\
  and\ \bibinfo {author} {\bibfnamefont {L.~B.}\ \bibnamefont {Ioffe}},\
  }\bibfield  {title} {\bibinfo {title} {Microscopic model of quantum butterfly
  effect: {Out-of-time-order} correlators and traveling combustion waves},\
  }\href {https://doi.org/https://doi.org/10.1016/j.aop.2016.09.006} {\bibfield
   {journal} {\bibinfo  {journal} {Ann. Phys. (N.Y.)}\ }\textbf {\bibinfo
  {volume} {375}},\ \bibinfo {pages} {378} (\bibinfo {year}
  {2016})}\BibitemShut {NoStop}%
\bibitem [{\citenamefont {Zhou}\ \emph {et~al.}(2023)\citenamefont {Zhou},
  \citenamefont {Guo}, \citenamefont {Xu}, \citenamefont {Chen},\ and\
  \citenamefont {Swingle}}]{Zhou2023}%
  \BibitemOpen
  \bibfield  {author} {\bibinfo {author} {\bibfnamefont {T.}~\bibnamefont
  {Zhou}}, \bibinfo {author} {\bibfnamefont {A.}~\bibnamefont {Guo}}, \bibinfo
  {author} {\bibfnamefont {S.}~\bibnamefont {Xu}}, \bibinfo {author}
  {\bibfnamefont {X.}~\bibnamefont {Chen}},\ and\ \bibinfo {author}
  {\bibfnamefont {B.}~\bibnamefont {Swingle}},\ }\bibfield  {title} {\bibinfo
  {title} {Hydrodynamic theory of scrambling in chaotic long-range interacting
  systems},\ }\href {https://doi.org/10.1103/PhysRevB.107.014201} {\bibfield
  {journal} {\bibinfo  {journal} {Phys. Rev. B}\ }\textbf {\bibinfo {volume}
  {107}},\ \bibinfo {pages} {014201} (\bibinfo {year} {2023})}\BibitemShut
  {NoStop}%
\bibitem [{\citenamefont {Nosov}\ \emph {et~al.}(2023)\citenamefont {Nosov},
  \citenamefont {Shapiro}, \citenamefont {Goldstein},\ and\ \citenamefont
  {Burmistrov}}]{Nosov2023}%
  \BibitemOpen
  \bibfield  {author} {\bibinfo {author} {\bibfnamefont {P.~A.}\ \bibnamefont
  {Nosov}}, \bibinfo {author} {\bibfnamefont {D.~S.}\ \bibnamefont {Shapiro}},
  \bibinfo {author} {\bibfnamefont {M.}~\bibnamefont {Goldstein}},\ and\
  \bibinfo {author} {\bibfnamefont {I.~S.}\ \bibnamefont {Burmistrov}},\
  }\bibfield  {title} {\bibinfo {title} {Reaction-diffusive dynamics of
  number-conserving dissipative quantum state preparation},\ }\href
  {https://doi.org/10.1103/PhysRevB.107.174312} {\bibfield  {journal} {\bibinfo
   {journal} {Phys. Rev. B}\ }\textbf {\bibinfo {volume} {107}},\ \bibinfo
  {pages} {174312} (\bibinfo {year} {2023})}\BibitemShut {NoStop}%
\bibitem [{\citenamefont {Lyublinskaya}\ \emph {et~al.}(2025)\citenamefont
  {Lyublinskaya}, \citenamefont {Nosov},\ and\ \citenamefont
  {Burmistrov}}]{Lyublinskaya2025}%
  \BibitemOpen
  \bibfield  {author} {\bibinfo {author} {\bibfnamefont {A.~A.}\ \bibnamefont
  {Lyublinskaya}}, \bibinfo {author} {\bibfnamefont {P.~A.}\ \bibnamefont
  {Nosov}},\ and\ \bibinfo {author} {\bibfnamefont {I.}~\bibnamefont
  {Burmistrov}},\ }\bibfield  {title} {\bibinfo {title} {Instability of the
  engineered dark state in two-band fermions under number-conserving
  dissipative dynamics},\ }\href {https://doi.org/10.1103/PhysRevB.111.094302}
  {\bibfield  {journal} {\bibinfo  {journal} {Physical Review B}\ }\textbf
  {\bibinfo {volume} {111}},\ \bibinfo {pages} {094302} (\bibinfo {year}
  {2025})}\BibitemShut {NoStop}%
\bibitem [{\citenamefont {Tonielli}\ \emph {et~al.}(2020)\citenamefont
  {Tonielli}, \citenamefont {Budich}, \citenamefont {Altland},\ and\
  \citenamefont {Diehl}}]{Tonielli2020}%
  \BibitemOpen
  \bibfield  {author} {\bibinfo {author} {\bibfnamefont {F.}~\bibnamefont
  {Tonielli}}, \bibinfo {author} {\bibfnamefont {J.~C.}\ \bibnamefont
  {Budich}}, \bibinfo {author} {\bibfnamefont {A.}~\bibnamefont {Altland}},\
  and\ \bibinfo {author} {\bibfnamefont {S.}~\bibnamefont {Diehl}},\ }\bibfield
   {title} {\bibinfo {title} {Topological field theory far from equilibrium},\
  }\href {https://doi.org/10.1103/PhysRevLett.124.240404} {\bibfield  {journal}
  {\bibinfo  {journal} {Phys. Rev. Lett.}\ }\textbf {\bibinfo {volume} {124}},\
  \bibinfo {pages} {240404} (\bibinfo {year} {2020})}\BibitemShut {NoStop}%
\bibitem [{\citenamefont {Lyublinskaya}\ and\ \citenamefont
  {Burmistrov}(2023)}]{Lyublinskaya2023}%
  \BibitemOpen
  \bibfield  {author} {\bibinfo {author} {\bibfnamefont {A.~A.}\ \bibnamefont
  {Lyublinskaya}}\ and\ \bibinfo {author} {\bibfnamefont {I.~S.}\ \bibnamefont
  {Burmistrov}},\ }\bibfield  {title} {\bibinfo {title} {Diffusive modes of
  two-band fermions under number-conserving dissipative dynaics},\ }\href
  {https://doi.org/10.1134/s0021364023602713} {\bibfield  {journal} {\bibinfo
  {journal} {JETP Letters}\ }\textbf {\bibinfo {volume} {118}},\ \bibinfo
  {pages} {524–530} (\bibinfo {year} {2023})}\BibitemShut {NoStop}%
\bibitem [{\citenamefont {Sieberer}\ \emph {et~al.}(2025)\citenamefont
  {Sieberer}, \citenamefont {Buchhold}, \citenamefont {Marino},\ and\
  \citenamefont {Diehl}}]{Sieberer2025}%
  \BibitemOpen
  \bibfield  {author} {\bibinfo {author} {\bibfnamefont {L.~M.}\ \bibnamefont
  {Sieberer}}, \bibinfo {author} {\bibfnamefont {M.}~\bibnamefont {Buchhold}},
  \bibinfo {author} {\bibfnamefont {J.}~\bibnamefont {Marino}},\ and\ \bibinfo
  {author} {\bibfnamefont {S.}~\bibnamefont {Diehl}},\ }\bibfield  {title}
  {\bibinfo {title} {Universality in driven open quantum matter},\ }\href
  {https://doi.org/10.1103/RevModPhys.97.025004} {\bibfield  {journal}
  {\bibinfo  {journal} {Reviews of Modern Physics}\ }\textbf {\bibinfo {volume}
  {97}},\ \bibinfo {pages} {025004} (\bibinfo {year} {2025})}\BibitemShut
  {NoStop}%
\bibitem [{\citenamefont {Yakupov}\ \emph {et~al.}(2019)\citenamefont
  {Yakupov}, \citenamefont {Polezhaev}, \citenamefont {Gubernov},\ and\
  \citenamefont {Miroshnichenko}}]{Yakupov2019}%
  \BibitemOpen
  \bibfield  {author} {\bibinfo {author} {\bibfnamefont {E.~O.}\ \bibnamefont
  {Yakupov}}, \bibinfo {author} {\bibfnamefont {A.}~\bibnamefont {Polezhaev}},
  \bibinfo {author} {\bibfnamefont {V.}~\bibnamefont {Gubernov}},\ and\
  \bibinfo {author} {\bibfnamefont {T.}~\bibnamefont {Miroshnichenko}},\
  }\bibfield  {title} {\bibinfo {title} {Investigation of the mechanism of
  emergence of autowave structures at the reaction front},\ }\href
  {https://doi.org/10.1103/PhysRevE.99.042215} {\bibfield  {journal} {\bibinfo
  {journal} {Physical Review E}\ }\textbf {\bibinfo {volume} {99}},\ \bibinfo
  {pages} {042215} (\bibinfo {year} {2019})}\BibitemShut {NoStop}%
\bibitem [{\citenamefont {Estavoyer}\ and\ \citenamefont
  {Lepoutre}(2024)}]{Estavoyer2024}%
  \BibitemOpen
  \bibfield  {author} {\bibinfo {author} {\bibfnamefont {M.}~\bibnamefont
  {Estavoyer}}\ and\ \bibinfo {author} {\bibfnamefont {T.}~\bibnamefont
  {Lepoutre}},\ }\bibfield  {title} {\bibinfo {title} {Travelling waves for a
  fast reaction limit of a discrete coagulation--fragmentation model with
  diffusion and proliferation},\ }\href
  {https://doi.org/10.1007/s00285-024-02099-4} {\bibfield  {journal} {\bibinfo
  {journal} {Journal of Mathematical Biology}\ }\textbf {\bibinfo {volume}
  {89}},\ \bibinfo {pages} {2} (\bibinfo {year} {2024})}\BibitemShut {NoStop}%
\bibitem [{\citenamefont {Berestycki}\ \emph {et~al.}(2025)\citenamefont
  {Berestycki}, \citenamefont {Rossi},\ and\ \citenamefont
  {Tellini}}]{Berestycki2025}%
  \BibitemOpen
  \bibfield  {author} {\bibinfo {author} {\bibfnamefont {H.}~\bibnamefont
  {Berestycki}}, \bibinfo {author} {\bibfnamefont {L.}~\bibnamefont {Rossi}},\
  and\ \bibinfo {author} {\bibfnamefont {A.}~\bibnamefont {Tellini}},\
  }\bibfield  {title} {\bibinfo {title} {Coupled reaction--diffusion equations
  on adjacent domains},\ }\bibfield  {journal} {\bibinfo  {journal}
  {Communications in Contemporary Mathematics}\ }\href
  {https://doi.org/10.1142/s0219199725500567} {10.1142/s0219199725500567}
  (\bibinfo {year} {2025})\BibitemShut {NoStop}%
\bibitem [{\citenamefont {Venegas-Ortiz}\ \emph {et~al.}(2014)\citenamefont
  {Venegas-Ortiz}, \citenamefont {Allen},\ and\ \citenamefont
  {Evans}}]{Ortiz2014}%
  \BibitemOpen
  \bibfield  {author} {\bibinfo {author} {\bibfnamefont {J.}~\bibnamefont
  {Venegas-Ortiz}}, \bibinfo {author} {\bibfnamefont {R.~J.}\ \bibnamefont
  {Allen}},\ and\ \bibinfo {author} {\bibfnamefont {M.~R.}\ \bibnamefont
  {Evans}},\ }\bibfield  {title} {\bibinfo {title} {Speed of invasion of an
  expanding population by a horizontally transmitted trait},\ }\href@noop {}
  {\bibfield  {journal} {\bibinfo  {journal} {Genetics}\ }\textbf {\bibinfo
  {volume} {196}},\ \bibinfo {pages} {497} (\bibinfo {year}
  {2014})}\BibitemShut {NoStop}%
\bibitem [{\citenamefont {Fu}\ and\ \citenamefont {Tsai}(2015)}]{Fu2015}%
  \BibitemOpen
  \bibfield  {author} {\bibinfo {author} {\bibfnamefont {S.-C.}\ \bibnamefont
  {Fu}}\ and\ \bibinfo {author} {\bibfnamefont {J.-C.}\ \bibnamefont {Tsai}},\
  }\bibfield  {title} {\bibinfo {title} {Wave propagation in predator--prey
  systems},\ }\href {https://doi.org/10.1088/0951-7715/28/12/4389} {\bibfield
  {journal} {\bibinfo  {journal} {Nonlinearity}\ }\textbf {\bibinfo {volume}
  {28}},\ \bibinfo {pages} {4389} (\bibinfo {year} {2015})}\BibitemShut
  {NoStop}%
\bibitem [{\citenamefont {Yadav}\ \emph {et~al.}(2025)\citenamefont {Yadav},
  \citenamefont {Sen},\ and\ \citenamefont {Pal}}]{Yadav2025aa}%
  \BibitemOpen
  \bibfield  {author} {\bibinfo {author} {\bibfnamefont {R.}~\bibnamefont
  {Yadav}}, \bibinfo {author} {\bibfnamefont {M.}~\bibnamefont {Sen}},\ and\
  \bibinfo {author} {\bibfnamefont {S.}~\bibnamefont {Pal}},\ }\bibfield
  {title} {\bibinfo {title} {Impact of diffusion-driven instability on
  traveling wave solutions},\ }\href
  {https://doi.org/10.1103/PhysRevE.111.024202} {\bibfield  {journal} {\bibinfo
   {journal} {Physical Review E}\ }\textbf {\bibinfo {volume} {111}},\ \bibinfo
  {pages} {024202} (\bibinfo {year} {2025})}\BibitemShut {NoStop}%
\bibitem [{\citenamefont {Holzer}(2014)}]{HOLZER20141}%
  \BibitemOpen
  \bibfield  {author} {\bibinfo {author} {\bibfnamefont {M.}~\bibnamefont
  {Holzer}},\ }\bibfield  {title} {\bibinfo {title} {Anomalous spreading in a
  system of coupled fisher--kpp equations},\ }\href
  {https://doi.org/https://doi.org/10.1016/j.physd.2013.12.003} {\bibfield
  {journal} {\bibinfo  {journal} {Physica D: Nonlinear Phenomena}\ }\textbf
  {\bibinfo {volume} {270}},\ \bibinfo {pages} {1} (\bibinfo {year}
  {2014})}\BibitemShut {NoStop}%
\bibitem [{\citenamefont {Faye}\ and\ \citenamefont {Holzer}(2019)}]{Faye2019}%
  \BibitemOpen
  \bibfield  {author} {\bibinfo {author} {\bibfnamefont {G.}~\bibnamefont
  {Faye}}\ and\ \bibinfo {author} {\bibfnamefont {M.}~\bibnamefont {Holzer}},\
  }\bibfield  {title} {\bibinfo {title} {Bifurcation to locked fronts in two
  component reaction--diffusion systems},\ }\href@noop {} {\bibfield  {journal}
  {\bibinfo  {journal} {AIHPC}\ }\textbf {\bibinfo {volume} {36}},\ \bibinfo
  {pages} {545} (\bibinfo {year} {2019})}\BibitemShut {NoStop}%
\bibitem [{\citenamefont {Keldysh}(1960)}]{Keldysh1960}%
  \BibitemOpen
  \bibfield  {author} {\bibinfo {author} {\bibfnamefont {L.~V.}\ \bibnamefont
  {Keldysh}},\ }\bibfield  {title} {\bibinfo {title} {{Kinetic theory of impact
  ionization in semiconductors}},\ }\href
  {http://www.jetp.ras.ru/cgi-bin/e/index/r/37/3/p713?a=list} {\bibfield
  {journal} {\bibinfo  {journal} {Sov. Phys. JETP}\ }\textbf {\bibinfo {volume}
  {10}},\ \bibinfo {pages} {509} (\bibinfo {year} {1960})}\BibitemShut
  {NoStop}%
\bibitem [{\citenamefont {Abakumov}\ \emph {et~al.}(1991)\citenamefont
  {Abakumov}, \citenamefont {Perel},\ and\ \citenamefont
  {Yassievich}}]{Abakumov-book}%
  \BibitemOpen
  \bibfield  {author} {\bibinfo {author} {\bibfnamefont {V.~N.}\ \bibnamefont
  {Abakumov}}, \bibinfo {author} {\bibfnamefont {V.~I.}\ \bibnamefont
  {Perel}},\ and\ \bibinfo {author} {\bibfnamefont {I.~N.}\ \bibnamefont
  {Yassievich}},\ }\href@noop {} {\emph {\bibinfo {title} {Nonradiative
  recombination in semiconductors}}}\ (\bibinfo  {publisher} {North-Holland},\
  \bibinfo {year} {1991})\BibitemShut {NoStop}%
\bibitem [{\citenamefont {Ablowitz}\ and\ \citenamefont
  {Zeppetella}(1979)}]{Ablowitz1979}%
  \BibitemOpen
  \bibfield  {author} {\bibinfo {author} {\bibfnamefont {M.~J.}\ \bibnamefont
  {Ablowitz}}\ and\ \bibinfo {author} {\bibfnamefont {A.}~\bibnamefont
  {Zeppetella}},\ }\bibfield  {title} {\bibinfo {title} {Explicit solutions of
  fisher's equation for a special wave speed},\ }\href
  {https://doi.org/10.1007/BF02462380} {\bibfield  {journal} {\bibinfo
  {journal} {Bulletin of Mathematical Biology}\ }\textbf {\bibinfo {volume}
  {41}},\ \bibinfo {pages} {835} (\bibinfo {year} {1979})}\BibitemShut
  {NoStop}%
\bibitem [{\citenamefont {Kamenskii}\ and\ \citenamefont
  {Manakov}(1987)}]{Manakov1987}%
  \BibitemOpen
  \bibfield  {author} {\bibinfo {author} {\bibfnamefont {V.}~\bibnamefont
  {Kamenskii}}\ and\ \bibinfo {author} {\bibfnamefont {S.}~\bibnamefont
  {Manakov}},\ }\bibfield  {title} {\bibinfo {title} {Formation of stability
  regions from unstable states in dissipative nonlinear systems},\ }\href@noop
  {} {\bibfield  {journal} {\bibinfo  {journal} {JETP Lett.}\ }\textbf
  {\bibinfo {volume} {45}},\ \bibinfo {pages} {638} (\bibinfo {year}
  {1987})}\BibitemShut {NoStop}%
\end{thebibliography}%
%\bibliography{FKPP-ref}

\end{document}